\documentclass[twocolumn,floatfix,superscriptaddress,a4paper,showpacs,showkeys,nofootinbib,reprint,prl]{revtex4-1}
\usepackage{epsfig}
\usepackage{latexsym}
\usepackage{xspace}
\usepackage[colorlinks=true,linktocpage=true,linkcolor=blue,citecolor=blue,allcolors=blue]{hyperref}
\usepackage{url}
\usepackage[utf8]{inputenc}
\usepackage{indentfirst}
\usepackage{enumerate}
\usepackage{color}

\usepackage[caption=false,position=top]{subfig}

\usepackage{amsmath}
\usepackage{amssymb}

\usepackage[english]{babel}
\usepackage{url}

\AtBeginDocument{\renewcommand{\natexlab}[1]{#1}}

\newcommand{\mean}[1]{\langle #1 \rangle}

\newcommand{\scvar}[1]{R_{#1-\bar{#1}}}

\newcommand{\scvarsum}[1]{R_{#1+\bar{#1}}}

\newcommand{\eq}[1]{\begin{align} #1 \end{align}}

\begin{document}

\title{
Constraining baryon annihilation in the hadronic phase of heavy-ion collisions via event-by-event fluctuations
}

\author{Oleh Savchuk}
\affiliation{Frankfurt Institute for Advanced Studies, Giersch Science Center, Ruth-Moufang-Str. 1, D-60438 Frankfurt am Main, Germany}

\author{Volodymyr Vovchenko}
\affiliation{Nuclear Science Division, Lawrence Berkeley National Laboratory, 1 Cyclotron Road, Berkeley, CA 94720, USA}

\author{Volker Koch}
\affiliation{Nuclear Science Division, Lawrence Berkeley National Laboratory, 1 Cyclotron Road, Berkeley, CA 94720, USA}

\author{Jan Steinheimer}
\affiliation{Frankfurt Institute for Advanced Studies, Giersch Science Center, Ruth-Moufang-Str. 1, D-60438 Frankfurt am Main, Germany}

\author{Horst Stoecker}
\affiliation{Frankfurt Institute for Advanced Studies, Giersch Science Center, Ruth-Moufang-Str. 1, D-60438 Frankfurt am Main, Germany}
\affiliation{Institut f\"{u}r Theoretische Physik, Goethe Universit\"{a}t Frankfurt, Max-von-Laue-Str. 1, D-60438 Frankfurt am Main, Germany}
\affiliation{
GSI Helmholtzzentrum f\"ur Schwerionenforschung GmbH, Planckstr. 1, D-64291 Darmstadt, Germany}

\begin{abstract}
We point out that the variance of net-baryon distribution normalized by the Skellam distribution baseline, $\kappa_2[B-\bar{B}]/\mean{B+\bar{B}}$, is sensitive to the possible modification of (anti)baryon yields due to $B\bar{B}$ annihilation in the hadronic phase.
The corresponding measurements can thus place stringent limits on the magnitude of the $B\bar{B}$ annihilation
and its inverse reaction.
We perform Monte Carlo simulations of the hadronic phase in Pb-Pb collisions at the LHC via the recently developed subensemble sampler + UrQMD afterburner and show that the effect survives in net-proton fluctuations, which are directly accessible experimentally.
The available experimental data of the ALICE Collaboration on net-proton fluctuations disfavors a notable suppression of (anti)baryon yields in $B\bar{B}$ annihilations predicted by the present version of UrQMD if only global baryon conservation is incorporated.
On the other hand, the annihilations improve the data description when local baryon conservation is imposed.
The two effects can be disentangled by measuring $\kappa_2[B+\bar{B}]/\mean{B+\bar{B}}$, which at the LHC is notably suppressed by annihilations but virtually unaffected by baryon number conservation.

\end{abstract}

\maketitle

\paragraph{\bf Introduction.}

Baryon-antibaryon annihilation is among the most important reactions during the hadronic phase of heavy-ion collisions. 
With a very large cross section ($>$60~mb) it has a potentially strong effect on the hadrochemical composition in the final state~\cite{Bass:2000ib}.
The hadronic transport models like UrQMD~\cite{Bass:1998ca,Bleicher:1999xi} or SMASH~\cite{Weil:2016zrk} predict sizable suppression of (anti)baryons yields due to the annihilations~\cite{Becattini:2012sq,Steinheimer:2012rd,Becattini:2012xb,Oliinychenko:2018ugs}.
This provides a possible resolution to the thermal proton yield anomaly at the LHC~\cite{Abelev:2013vea}.
However, while the transport models do incorporate the annihilation reactions like $B+\bar{B} \to n \,\pi$~\cite{Cassing:1999es, Lin:2004en}, a proper implementation of all the relevant reverse regeneration reactions $n \pi \to B + \bar{B}$ remains challenging~\cite{Garcia-Montero:2021haa}.
% and require experimental verification.
These reactions can mitigate the effect of annihilations to some extent, if not negate it completely~\cite{Rapp:2000gy,Pan:2012ne,Pan:2014caa,Seifert:2018bwl}.
Furthermore, possibilities to explain the thermal proton yield anomaly have been suggested based on reevaluating the chemical equilibrium proton abundances at the conventional chemical freeze-out~\cite{Alba:2016hwx,Vovchenko:2018fmh,Andronic:2018qqt}.

In this work we present a new way to constrain and quantify the effect of $B\bar{B}$ annihilations on (anti)baryon abundances by utilizing measurements of event-by-event fluctuations.
Cumulants of the (net-)(anti)proton distributions have recently been measured in various experiments, including the ALICE Collaboration at the LHC~\cite{Acharya:2019izy}, the STAR Collaboration at RHIC~\cite{Adam:2020kzk,Abdallah:2021fzj}, and the HADES Collaboration at GSI~\cite{Adamczewski-Musch:2020slf}.
These measurements have primarily been motivated to probe the phase structure of QCD, in particular in the hunt for the hypothetical QCD critical point~\cite{Stephanov:1999zu,Bzdak:2019pkr}.
Here, we point out that specific combinations of first and second moments of baryon (proton) numbers are sensitive to $B\bar{B}$ annihilations.

\paragraph{\bf Annihilations and fluctuations.}

Let us denote by $\gamma_{B(\bar{B})}$ the modification factors of the mean yields of (anti)baryons during the hadronic phase for a particular collision energy and centrality, i.e.
\eq{
\mean{N_{B(\bar{B})}^{\rm fin}} = \gamma_{B(\bar{B})} \, \mean{N_{B(\bar{B})}^{\rm hyd}}~.
}
$\mean{N_{B(\bar{B})}^{\rm fin}}$ is the final mean yield of (anti)baryons and $\mean{N_{B(\bar{B})}^{\rm hyd}}$ is the mean yield at the beginning of the hadronic phase, i.e. at the end of the hydrodynamic evolution.
$\gamma_{B(\bar{B})} = 1$ corresponds to a vanishing net effect of baryon annihilations and regenerations whereas $\gamma_{B(\bar{B})} < 1$ corresponds to a suppression of the yields as observed in transport models.

Instead of the mean, let us consider now the variance $\kappa_2[B-\bar{B}]$ of net baryon distribution.
Since the net baryon number is unchanged in $B\bar{B}$ annihilations, or in any other QCD process for that matter,
 $\kappa_2[B-\bar{B}]$ is unaffected by the hadronic phase evolution, as long as the diffusion of baryons in and out of the acceptance can be neglected:
 \eq{
 \kappa_2^{\rm fin}[B-\bar{B}] = \kappa_2^{\rm hyd}[B-\bar{B}]
 }
 
To obtain an intensive~(volume-independent) measure of the net baryon number fluctuations it is convenient to normalize $\kappa_2^{\rm fin}[B-\bar{B}]$ by the mean number of baryons and antibaryons, $\mean{N_{B} + N_{\bar{B}}}$:

\eq{\label{eq:kappa2}
\scvar{B}^{\rm fin} \equiv \frac{\kappa_2^{\rm fin}[B-\bar{B}]}{\mean{N_{B}^{\rm fin} + N_{\bar{B}}^{\rm fin}}} =\frac{\kappa_2^{\rm hyd}[B-\bar{B}]}{\mean{\gamma_B\, N_{B}^{\rm hyd} + \gamma_{\bar{B}} \, N_{\bar{B}}^{\rm hyd}}}.
}

It follows from Eq.~\eqref{eq:kappa2} that any suppression of the baryon yields in the hadronic phase~($\gamma_{B(\bar{B})} < 1$) leads to an enhancement of the normalized net baryon variance.
Measurement of such an enhancement is the key idea here to constrain the annihilation.

\paragraph{\bf Fluctuations at the LHC.}

To isolate the effect of annihilation on $\scvar{B}$ other physical mechanisms affecting this quantity must be well controlled.
In the limit of uncorrelated baryon production at particlization, as is the case for the ideal hadron resonance gas~(HRG) model in the grand-canonical ensemble, the quantity 
$\scvar{B}^{\rm hyd} = \kappa_2^{\rm hyd}[B-\bar{B}] / \mean{N_{B}^{\rm hyd} + N_{\bar{B}}^{\rm hyd}}$
would be equal to unity.
However, it is known from first-principle lattice calculations that cumulants of the baryon number distribution in QCD deviate from this baseline at temperatures $T \sim 150 - 160$~MeV relevant for particlization~\cite{Bazavov:2017dus,Borsanyi:2018grb}.
Furthermore, baryon number fluctuations in heavy-ion collisions are affected by exact conservation of baryon number~\cite{Bzdak:2012an}.
These two issues have recently been addressed in Ref.~\cite{Vovchenko:2020kwg} via a generalized Cooper-Frye particlization routine called \textit{subensemble sampler}.
In the following we will employ this sampler to evaluate the effect of $B\bar{B}$ annihilations on experimentally observable event-by-event fluctuations.

We restrict our calculations to 2.76~TeV central Pb-Pb collisions, where measurements of net-proton fluctuations done by the ALICE Collaboration have recently become available~\cite{Acharya:2019izy}.
The general arguments are valid also for the lower collision energies probed by other experiments but the conditions created at the LHC allow several simplifications to make the interpretation and analysis clearer.
At this high beam energy, the effect of baryon annihilation will be most obvious since baryons and anti-baryons are created in equal amounts.
The measurements of the net-particle number variances at the LHC have an additional advantage, since they are less affected by volume fluctuations~\cite{Skokov:2012ds,Braun-Munzinger:2016yjz}. 
The subensemble sampler of Ref.~\cite{Vovchenko:2020kwg} is used to generate events consisting of hadrons and resonances at particlization, which is assumed to take place at $T = 160$~MeV and $\mu_B = 0$.
The procedure incorporates baryon excluded volume effects, matched to lattice QCD data on baryon susceptibilities~\cite{Vovchenko:2017xad}, as well as exact global baryon conservation~(see Ref.~\cite{Vovchenko:2020kwg} for details).
Here we neglect the exact conservation of electric charge and strangeness which at the LHC energies where shown to have only small effect on baryon and proton number fluctuations~\cite{Vovchenko:2020kwg}.
The particlization hypersurface incorporates the collective flow~\cite{Siemens:1978pb,Stoecker:1981za}, taken from the longitudinally boost-invariant blast-wave model~\cite{Schnedermann:1993ws}, which corresponds to a cylinder $r_\perp < r_{\rm max}$ in the transverse plane at a constant value $\tau = \tau_0$ of the longitudinal proper time.
The transverse velocity is parameterized as $\beta_{\rm \perp} = \beta_s (r_{\perp} / r_{\rm max})^n$, with parameters $\beta_s = 0.8$, $r_{\rm max} = 10$~fm, and $n = 1$ taken from Ref.~\cite{Huovinen:2016xxq}.
The value of $\tau_0 = 11.6$~fm/$c$ is taken to reproduce the effective volume per unit of rapidity $dV/dy \sim 4000$~fm$^3$ at particlization, as suggested by thermal models~\cite{Andronic:2017pug,Vovchenko:2018fmh}.
We checked that with this choice of parameters the resulting hypersurface accurately reproduces the results of numerical fluid dynamic simulations at the LHC within the UrQMD+hybrid model~\cite{Petersen:2008dd}.
To account for the finite longitudinal extent of the system we impose a cut-off $|\eta_s| < 4.8$ on the longitudinal space-time rapidity, which, as shown in Ref.~\cite{Vovchenko:2020kwg}, in the boost-invariant scenario accurately reproduces the experimental estimates for the full acceptance charged particle multiplicity~\cite{Abbas:2013bpa}.
This is necessary to properly take into account exact global (or local) baryon number conservation.

The sampled hadrons and resonances in each event are then injected into the hadronic afterburner UrQMD.
We run UrQMD using two different configurations: 
\begin{enumerate}
    \item the standard configuration implementing $B\bar{B}$ annihilations and
    \item the second configuration where these reactions were switched off.
\end{enumerate}

This allows us to establish the two extreme cases: full annihilation without regeneration which maximizes the effect, and no annihilation at all.
We additionally consider a third scenario, where we neglect the afterburner stage but instead let all resonances produced at particlization decay immediately into final hadrons.
This scenario is labeled \textit{decays only} and has been studied in the original work, Ref.~\cite{Vovchenko:2020kwg}.
We sample 1.76, 2.4, and 2.4 million events for each of the three scenarios respectively.

First, it is verified that the mean proton multiplicities, as measured by the ALICE experiment, are reproduced.
Due to the annihilation, the mean number of baryons is suppressed by a factor $\gamma_{B(\bar{B})} = 0.84$, consistent with earlier results~\cite{Becattini:2016xct}.
The mean baryon multiplicity remains unchanged during the hadronic phase if annihilations are switched off.
The effect of the hadronic rescattering on the transverse momentum spectra of protons, is to increase the mean $p_T$ from $\mean{p_T} = 1.14$~GeV/$c$~(decays only) to $\mean{p_T} = 1.32$~GeV/$c$~(afterburner with annihilation) and $\mean{p_T} = 1.36$~GeV/$c$~(afterburner w/o  annihilation), improving the agreement with the ALICE data~($\mean{p_T} = 1.33 \pm 0.03$~GeV/$c$~\cite{Abelev:2013vea}). This means that the mean transverse momentum exhibits little sensitivity to $B\bar{B}$ annihilation. It is consistent with observations from earlier afterburner studies~\cite{Steinheimer:2017vju,Oliinychenko:2018ugs} where the change in the transverse momenta is mainly due to the abundant baryon+meson (pseudo-)elastic scatterings~\cite{Bleicher:2002dm}.

Note, that the blast-wave model used here for particlization is a simplified description compared to the state-of-the-art viscous hydrodynamic simulations of heavy-ion collisions.
Although a proper choice of blast-wave model parameters leads to hadron transverse momentum spectra and the effects of baryon annihilation that are comparable to that emerging from full hydro simulations~\cite{Steinheimer:2017vju}, it may be interesting to perform a similar analysis within full hydro simulations, including the effects of shear and bulk viscosities~\cite{Ryu:2015vwa}. However, this would require extending the particlization routine in such simulations and the generation of an extensive number of full hydro events to describe the event-by-event fluctuations of hadron multiplicities properly.
We would like to emphasize, however, that the purpose of the present paper is to demonstrate that cumulants provide useful constraint on baryon annihilation rather than to present a fully quantitative description.

\begin{figure}[t]
  \centering
  \includegraphics[width=.49\textwidth]{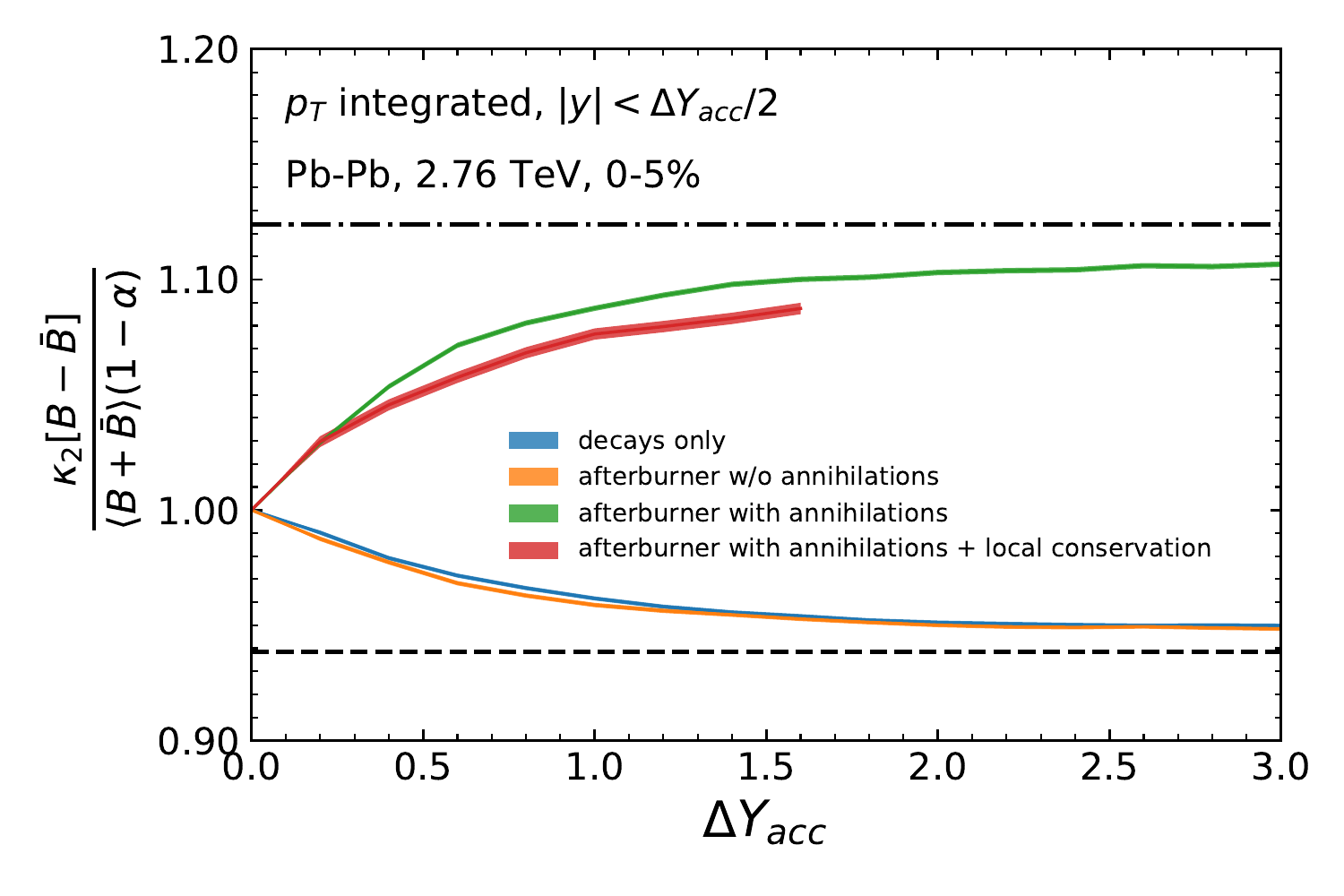}
  \caption{
  Rapidity acceptance dependence of net baryon $\scvar{B} = \kappa_2[B-\bar{B}]/\mean{B+\bar{B}}$ in 0-5\% central Pb-Pb collisions at the LHC, corrected  for baryon number conservation via the $1-\alpha$ factor.
  The results are obtained via the subensemble sampler particlization routine applied to the EV-HRG model and blast-wave hypersurface.
  The different bands correspond to different scenarios for the hadronic phase modeling.
  The red band corresponds to local baryon conservation in a rapidity range $\Delta Y_{\rm cons} = 3$.
  The horizontal bands depict the corresponding grand-canonical limits expected in the limit of large acceptance.
  }
  \label{fig:c2Bcor}
\end{figure}

Having fixed the mean multiplicity and ensured a realistic momentum distribution, we can now study the effects of annihilation in more detail. From our simulations we can calculate the second order cumulant $\kappa_2$ from the momentum integrated event-by-event multiplicities. 
As mentioned previously, this cumulant is affected by global baryon conservation.
The effect amounts to a multiplication by a suppression factor $(1-\alpha)$~\cite{Bleicher:2000ek}, where at the LHC $\alpha = \Delta Y_{\rm acc} /\Delta Y_{\rm tot}$~\cite{Vovchenko:2020kwg} with $\Delta Y_{\rm tot} = 9.6$ being the total rapidity width of the system with global conservation.
To subtract the effect of baryon conservation and maximize the effects of annihilations we thus analyze a scaled ratio $\scvar{B} \times \frac{1}{1-\alpha}$. 
This quantity is depiced in Fig.~\ref{fig:c2Bcor} as function of the rapidity acceptance window, $|y| < \Delta Y_{\rm acc} / 2$.

All presented curves tend to unity for $\Delta Y_{\rm acc} \to 0$. This ``poissonization'' is expected from a thermal system ~\cite{Ling:2015yau} in the $\Delta Y_{\rm acc} \to 0$ limit but goes away in the limit of large rapidity acceptance.

In the absence of any annihilations, the corrected $\scvar{B}$ exhibits essentially no sensitivity to the afterburner phase as the results from the immediate decays scenario virtually coincide with the afterburner without $B\bar{B}$ annihilation.
In this case the corrected scaled variance exhibits a mild decrease with $\Delta Y_{\rm acc}$, saturating at the grand canonical value of about $0.94$~(dashed line) at $\Delta Y_{\rm acc} \simeq 1.5$. This confirms our earlier assumption that baryon diffusion, excluding resonance decays, can be safely neglected at LHC energies in the estimation of the final state effect on the baryon number cumulants.

When $B\bar{B}$ annihilations are included, the scaled  variance exhibits the opposite behavior. As function of $\Delta Y_{\rm acc}$, it approaches a value of about 1.12~(dash-dotted line) at large $\Delta Y_{\rm acc}$.
This ratio at large $\Delta Y_{\rm acc}$ corresponds to what is expected from Eq.~\eqref{eq:kappa2} for a reduction of the mean baryon number 
\begin{equation}
\frac{\scvar{B}^{annih.}}{\scvar{B}^{no \, annih.}}=\frac{0.94}{1.12} \simeq 0.84 = \gamma_{B(\bar{B})}.
\end{equation}
This sensitivity of $\scvar{B}$ at $\Delta Y_{\rm acc} \gtrsim 1$ to the $B\bar{B}$-annihilations can be easily understood. While the mean multiplicity is reduced by $\gamma_{B(\bar{B})}$, the second order susceptibility $\kappa_2$ remains essentially unchanged.
It is straight forward to suggest, that this observable can be used to constrain the quantitative role of baryon annihilation and its back reaction during the hadronic rescattering.

\paragraph{\bf Comparison with experiment.}

\begin{figure}[t]
  \centering
  \includegraphics[width=.49\textwidth]{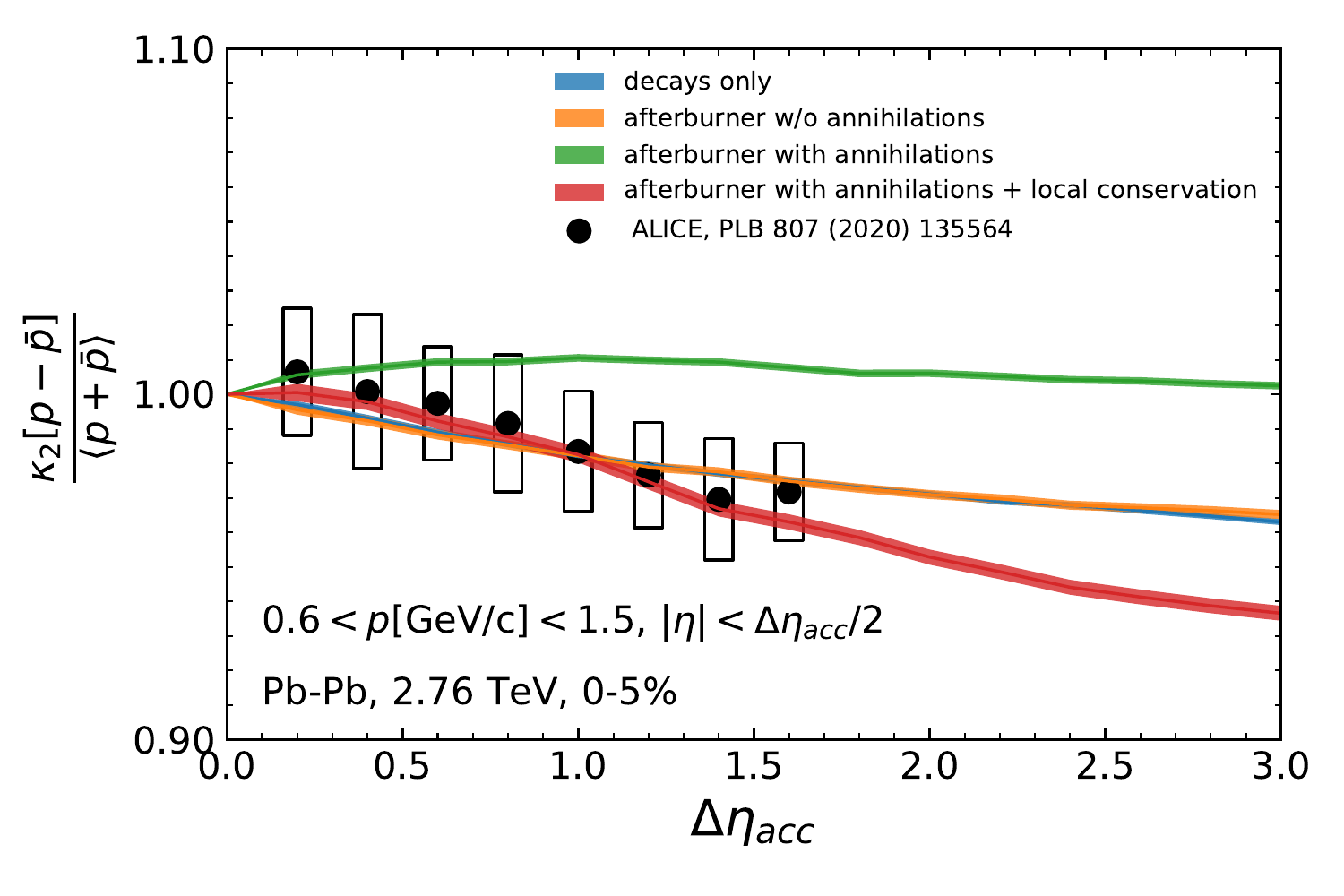}
  \caption{
  Pseudorapidity acceptance dependence of net proton $\scvar{p} =\kappa_2[p -\bar{p}]/\mean{p + \bar{p}}$ in 0-5\% central Pb-Pb collisions at the LHC. The bands have the same meaning as in Fig.~\ref{fig:c2Bcor}. The experimental data of the ALICE collaboration~\cite{Acharya:2019izy} are shown by the symbols with error bars.
  }
  \label{fig:c2pALICE}
\end{figure}

The simulation results on the net proton scaled variance $\scvar{p}$ are compared to the available experimental data from the ALICE Collaboration~\cite{Acharya:2019izy} in Fig.~\ref{fig:c2pALICE}. 
For this case we assume that protons can only be measured within the ALICE acceptance $0.6 < p_{T} < 1.5$~GeV/$c$ for various cuts in longitudinal pseudorapidity $|\eta| < \Delta \eta_{\rm acc}/2$.
Even though the effect of $B\bar{B}$ annihilation in such a reduced acceptance is not as evident as in the $p_T$ integrated net baryon fluctuations discussed above, the observable is still sensitive to annihilations. Note that, in contrast to Fig.~\ref{fig:c2Bcor}, the scaled variance shown in Fig.~\ref{fig:c2pALICE} is not corrected for baryon conservation.

We see that for the proton scaled variance, as for the baryons, the results from the decays only and afterburner w/o $B\bar{B}$ annihilation scenarios virtually coincide.
These scenarios are consistent with the experimental data within errors.
Since in this case the effects of global baryon conservation are not corrected, $\scvar{p}$ is reduced with increasing acceptance coverage.

When the $B\bar{B}$ annihilations are included~(afterburner with annihilations), the resulting acceptance dependence notably overestimates the ALICE data, for $\Delta \eta_{\rm acc} \gtrsim 1$.
Therefore, the available experimental data is inconsistent with a notable suppression of (anti)baryon yields in the hadronic phase as predicted by UrQMD in this setup.

\paragraph{\bf Local baryon conservation.}

However, there is another important property which can affect the scaled variance, namely that baryon number is conserved not only globally but also locally, i.e. in a limited rapidity window.
This possibility has been discussed in a previous publication~\cite{Acharya:2019izy}. There it was argued that the data on net-proton fluctuations in central Pb-Pb collisions would favor global over local baryon conservation. The argument for this conclusion was based on the observation that local conservation within $\Delta y_{\rm cons} \lesssim 5$ units of rapidity, would lead to a stronger suppression of $\scvar{p}$ than in the data.
However, the analysis in Ref.~\cite{Acharya:2019izy} did not consider the possible effects of $B\bar{B}$ annihilation.
As we have shown here, this effect leads to an enhancement of $\scvar{p}$ and thus could recover the agreement with the data. 
It is therefore essential to including annihilation for studying the effects of local conservation.

Several different ways of modeling local baryon conservation have been discussed in the literature~\cite{Castorina:2013mba,Vovchenko:2018fiy,Oliinychenko:2019zfk,Pruneau:2019baa,Vovchenko:2019kes,Braun-Munzinger:2019yxj,Altsybeev:2020qnd}.
We follow the approach of Refs.~\cite{Vovchenko:2018fiy,Vovchenko:2019kes} and incorporate the effect by reducing the space-time rapidity cut-off to $|\eta_s| < \Delta Y_{\rm cons} / 2$, where $\Delta Y_{\rm cons}$ is the longitudinal range in which baryon number is conserved.
This way we remove hadrons emitted at forward-backward rapidities~($|\eta_s| > \Delta Y_{\rm cons} / 2$) from contributing to the baryon~(proton) number fluctuations at midrapidity.
We select $\Delta Y_{\rm cons} = 3$, as it was earlier successfully used, in Ref.~\cite{Vovchenko:2019kes}, to describe the hadron yields in small systems at the LHC in the framework of the canonical statistical model.

The calculations including local baryon conservation for $\scvar{B}$~(0.4 million events) are shown in Figs.~\ref{fig:c2Bcor} and \ref{fig:c2pALICE} as red band. These calculations both include the $B\bar{B}$ annihilations during the hadronic phase. 
In Fig.~\ref{fig:c2Bcor} the correction for baryon conservation is performed via the $1-\alpha$ factor changed to $\alpha = \Delta Y_{\rm acc} / \Delta Y_{\rm cons}$, reflecting the local nature of baryon conservation.
It is seen that the corrected $\scvar{B}$ essentially coincides with the result of global conservation within the conservation radius $\Delta Y_{\rm acc} < 1.5$.
Thus, if the range of baryon conservation is known, the appropriately corrected $\scvar{B}$ can be used to constrain the baryon annihilation.

If the conservation range is not independently known, however, the picture is quite different. The combined effect of local baryon conservation and $B\bar{B}$ annihilations on $\scvar{p}$ in the ALICE acceptance is shown by the red band in Fig.~\ref{fig:c2pALICE}.
In this scenario, the calculation with local conservation and annihilation is in good agreement with the experimental data. 
In the absence of $B\bar{B}$ annihilations the data would be notably underestimated due to local conservation, as shown in Ref.~\cite{Acharya:2019izy}, however when both the local conservation and $B\bar{B}$ annihilation are implemented simultaneously, the agreement with the data is recovered.

\paragraph{\bf Distinguishing annihilation from local baryon conservation.}

The data presented by the ALICE collaboration currently does not allow us to distinguish global conservation without $B\bar{B}$ annihilations from local conservation with $B\bar{B}$ annihilations, although it can be argued that the shape of $\Delta \eta_{\rm acc}$ dependence is better reproduced by the latter scenario.
Additional analysis is required to answer this question more definitively and also possibly put quantitative constraints on the effect of annihilation and regeneration during the hadronic phase.

One option is to look into the centrality dependence of $\scvar{p}$.
The effect of the hadronic phase~(and thus $B\bar{B}$ annihilations) decreases for larger impact parameter, and can basically be neglected in peripheral collisions.
Experimental data on $\scvar{p}$~\cite{Acharya:2019izy} do indicate a centrality dependence: $\scvar{p}$ decreases from $0.972 \pm 0.015$ in 0-5\% central collisions to $0.935 \pm 0.011$ in 60-70\% central collisions.
The latter value was shown in Ref.~\cite{Vovchenko:2019kes} to be consistent with local baryon conservation with $\Delta Y_{\rm cons} = 3$ without $B\bar{B}$ annihilations.
If (local) baryon conservation is independent of centrality, for instance if it is determined by the quark-anti-quark creation in the early stage of the collision, the centrality dependence of the data favors the local conservation + $B\bar{B}$ annihilation scenario.
Additional support for this scenario can be found in the centrality dependence of the $p/\pi$ ratio, where the data show indications for suppression in central Pb-Pb collisions~\cite{Abelev:2013vea}, consistent with the effect of $B\bar{B}$ annihilations.

Besides these indications, another observable which is able to distinguish these two scenarios, based on experimental data, would be very useful.
To disentangle the local baryon conservation from $B\bar{B}$ annihilations more directly we propose to study an additional fluctuation measure.
In particular the scaled variance $\scvarsum{B} \equiv \kappa_2[B + \bar{B}]/\mean{N_B + N_{\bar{B}}}$~[or $\scvarsum{p}$ for protons] of the total baryon~(proton) + antibaryon~(antiproton) number can be used for this purpose.
This quantity is not sensitive to baryon conservation at the LHC because its correlator with the conserved net baryon number vanishes due to symmetry:
\eq{
\rm{cov}[B+\bar{B},B-\bar{B}] = \rm{cov}[B,B] - \rm{cov}[\bar{B},\bar{B}] \stackrel{\mean{B} = \mean{\bar{B}}}{=} 0~.
}

\begin{figure}[t]
  \centering
  \includegraphics[width=.49\textwidth]{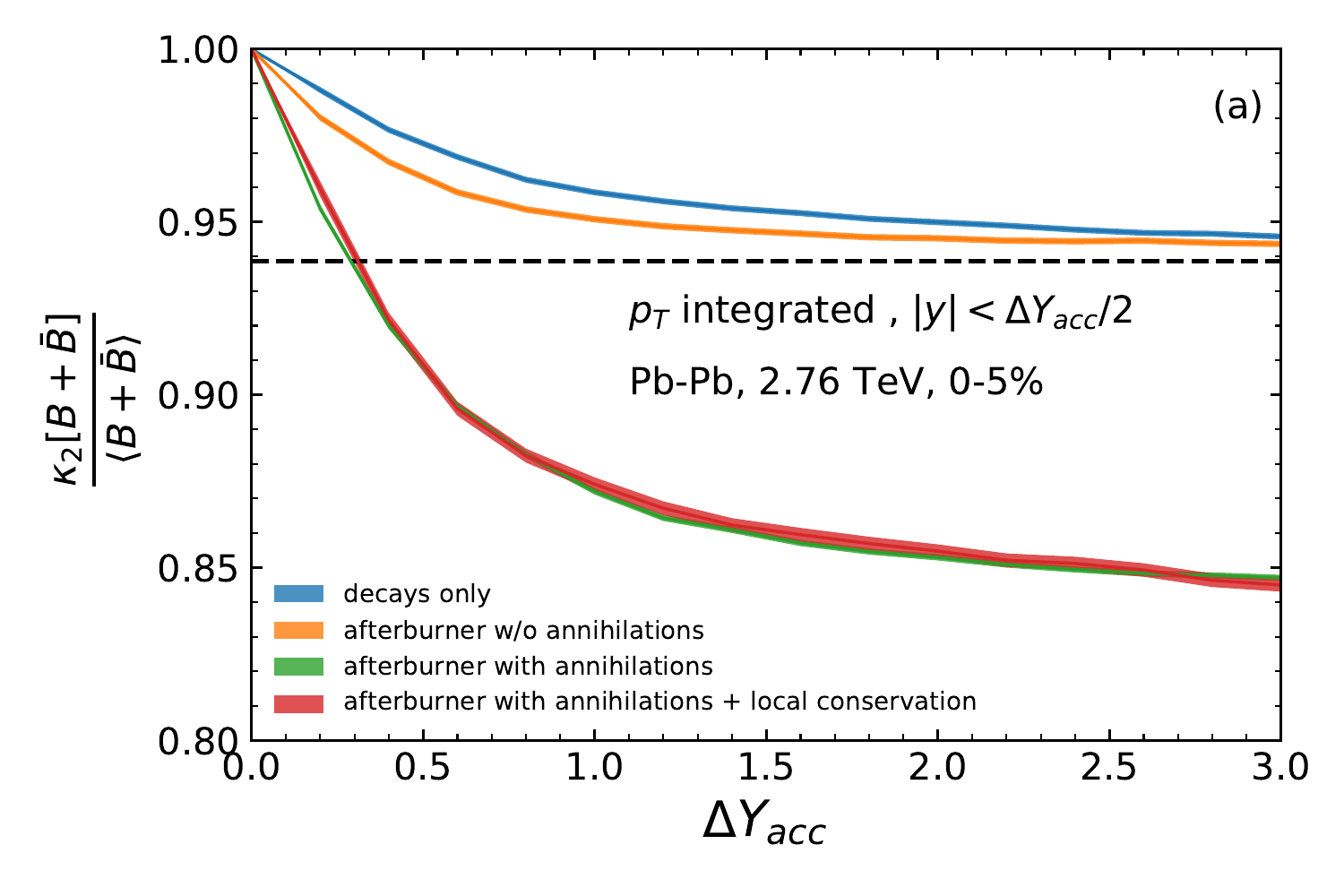}
   \includegraphics[width=.49\textwidth]{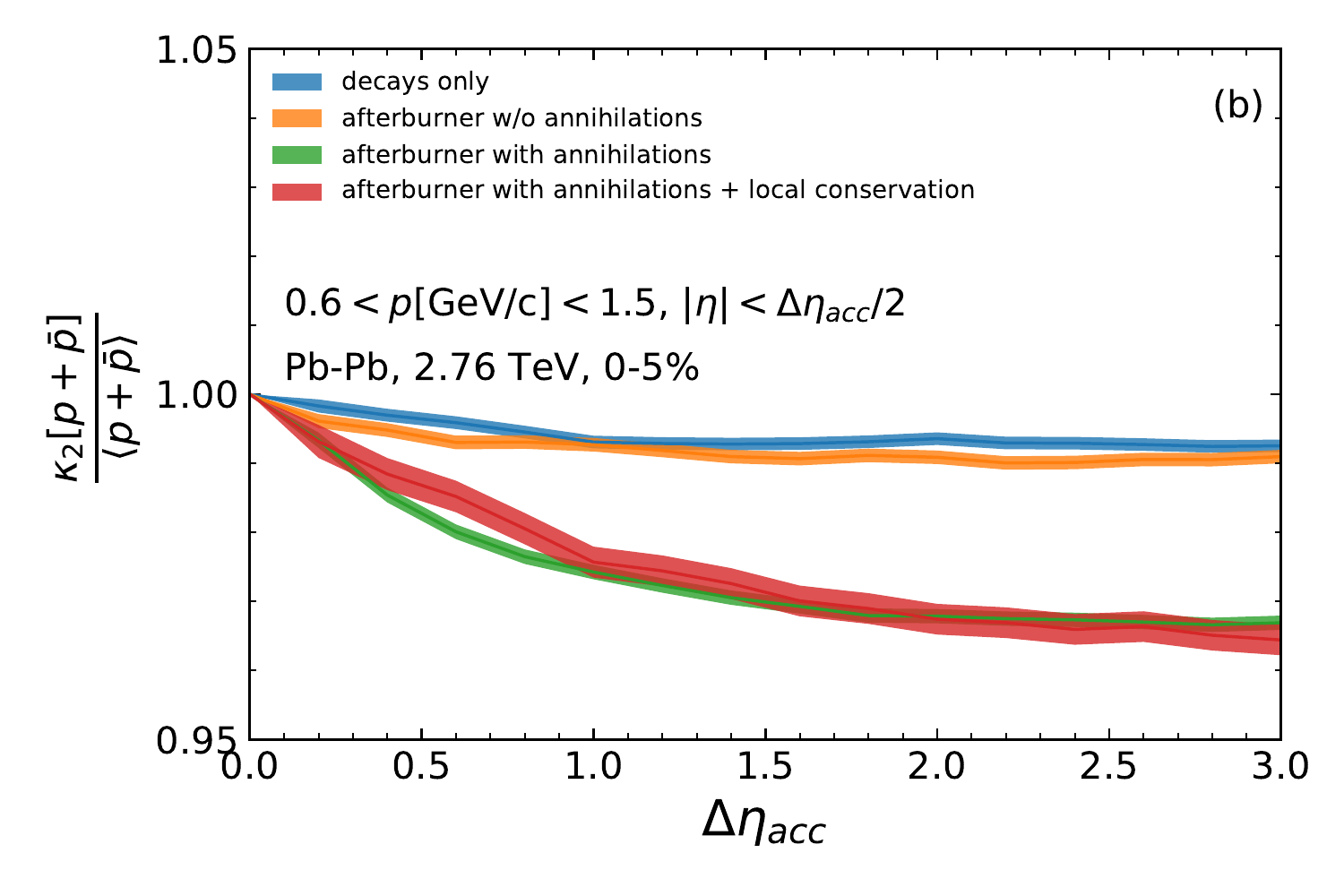}
  \caption{
  (a) Rapidity acceptance dependence of net baryon $\scvarsum{B} = \kappa_2[B+\bar{B}]/\mean{B+\bar{B}}$ in 0-5\% central Pb-Pb collisions at the LHC. The different bands have the same meaning as in Fig.~\ref{fig:c2Bcor}.
  (b) Same as Fig.~\ref{fig:c2pALICE} but for $\scvarsum{p}$.
  The calculations shown in this figure do not contain contributions from volume fluctuations.
  }
  \label{fig:c2sums}
\end{figure}

However, $\scvarsum{B}$ is sensitive to the annihilation and thus can be used to constrain this effect.
Figure~\ref{fig:c2sums} shows the results of calculations for (a) $\scvarsum{B}$ as function of the rapidity acceptance $\Delta Y_{\rm acc}$ and (b) $\scvarsum{p}$ as function of the pseudorapidity acceptance $\Delta \eta_{\rm acc}$ within the ALICE momentum acceptance $0.6 < p < 1.5$~GeV/$c$.
The numerical results explicitly illustrate that these two quantities are not sensitive to baryon number conservation: in the absence of $B\bar{B}$ annihilations $\scvarsum{B}$ approaches the grand-canonical value for large $\Delta Y_{\rm acc}$ whereas the calculations with the annihilations that incorporate either global or local baryon conservation yield identical results. 
The effect of $B\bar{B}$ annihilations is to suppress both quantities.
In particular, the suppression is notable for $\scvarsum{p}$ within the ALICE acceptance, thus the measurements can in principle be used to study $B\bar{B}$ annihilations independent of the (local) baryon conservation.

Note that, in contrast to the net charges, $\scvarsum{p}$ is significantly affected by volume fluctuations even at the LHC.
The calculations in Fig.~\ref{fig:c2sums} do not incorporate volume fluctuations, thus, for a meaningful comparison either the data have to be corrected for volume fluctuations or volume fluctuations included in the model calculation.
The data can be corrected using models for volume fluctuations, for example the Glauber Monte Carlo~\cite{Braun-Munzinger:2016yjz}.
We have checked that the errors for $\scvarsum{p}$ that can be derived from the published data on proton fluctuations~\cite{Acharya:2019izy}, as well as the systematic errors from performing the correction for volume fluctuations are presently too large to distinguish the difference between the annihilation scenarios shown in Fig.~\ref{fig:c2sums}.
However, this should be possible with upcoming high  precision data.
The event-by-event fluctuations presented here will also be useful in the ongoing efforts to properly implement  the regeneration reactions in hadronic afterburners.

Here we discussed the fluctuations either in $p_T$-integrated acceptance as function of rapidity cut or in the acceptance where the measurements of proton fluctuations have been performed by the ALICE Collaboration. The analysis can be further supplemented by systematic analysis of fluctuations as function of transverse momentum cuts, given the sizable imprint of annihilations on the $p_T$ spectra.

\paragraph{\bf Summary.}

We pointed out for the first time that measurements of event-by-event fluctuations in heavy-ion collisions can be used to quantify the effect of baryon annihilation in the hadronic phase.
The key observation is that the combined measurement of the scaled variance of the net and total proton + anti-proton number can be used to strictly constrain the role of baryon – anti-baryon annihilation during the hadronic phase of heavy ion collisions at the LHC. 
To this end, we studied two extreme scenarios where we either incorporate full annihilation without regeneration or do not include any annihilation at all.
Including only annihilation, without any regeneration, enhances appreciably the scaled variances of the net baryon and net proton distributions. A direct comparison with experimental data from the ALICE experiment leads to two possible scenarios which both equally well describe the data. Either, baryons are completely regenerated and baryon number is conserved only globally or, the mean baryon number is reduced due to annihilation which requires a more local conservation of the baryon number. 
In particular, if one neglects the effects of baryon pair regeneration in the hadronic phase, a conservation radius of $\Delta Y_{\rm cons} = 3$ units in rapidity is observed, while if some regeneration were included one would expect the $\Delta Y_{\rm cons}$ value that describes the data to be larger.
To distinguish the different possibilities we suggest a new observable: the scaled variance of the number of protons + anti-protons, which shows a clearly distinguished dependence on annihilation and local conservation. With a simultaneous observation of these two quantities one could finally experimentally establish the quantitative effect of annihilation during the hadronic phase.

\begin{acknowledgments}

\emph{Acknowledgments.} 
This work received support through the U.S. Department of Energy, 
Office of Science, Office of Nuclear Physics, under contract number 
DE-AC02-05CH11231231 and within the framework of the
Beam Energy Scan Theory (BEST) Topical Collaboration.
V.V. acknowledges the support through the
Feodor Lynen program of the Alexander von Humboldt
foundation. 
J.S. thank the Samson AG and the BMBF through the ErUM-Data project for funding and the DAAD-PPP program for support. H.St. acknowledges the Walter Greiner Gesellschaft zur F\"orderung der physikalischen Grundlagenforschung e.V. through the Judah M. Eisenberg Laureatus Chair at Goethe Universit\"at Frankfurt am Main.

\end{acknowledgments}

\bibliography{annihilations-via-fluctuations}

\end{document}